# Hydrodynamics of catheter biofilm formation

*Oscar Sotolongo-Costa[1,2], Manuel Arias-Zugasti[3], Daniel Rodríguez-Pérez[3], Sergio Martínez Escobar[1,4], Antonio Fernández-Barbero[1]*

[1]Complex Fluids Physics Group, Department of Applied Physics, University of Almeria, 04120-Almeria, Spain.

[2]Cátedra de Sistemas Complejos "Henri Poincaré". Universidad de La Habana, 10400, Cuba.

[3]Departamento de Física-Matemática y Fluidos, UNED, Madrid, Spain.

[4] Intensive Care Unit. University Hospital Torrecardenas. Public Andalucian Health Service, 04009-Almeria, Spain.

**Abstract**

A hydrodynamic model is proposed to describe one of the most critical problems in intensive medical care units: the formation of biofilms inside central venous catheters. The incorporation of approximate solutions for the flow-limited-diffusion equation leads to the conclusion that biofilms grow on the internal catheter wall due to the counter-stream diffusion of blood through a very thin layer close to the wall. This biological deposition is the first necessary step for the subsequent bacteria colonization.

**Introduction**

Growing of microorganism in association with a surface is frequently named biofilm formation. It needs a material substrate as demonstrated by the presence of non-infectious fibrin layers onto the catheters lumen. *(1-3)* As the sole possibility of substrate formation comes from the accumulation of bood proteins, it is of special importance the study of the mechanisms of catheter blood penetration (CBP).

From a clinical point of view, central venous catheterization use to cause undesired and critical complications, being one of the most frequent the bloodstream infection.(4, 5) This process is strongly related to the biological interactions between the catheter polymer surface and the bloodstream. These interactions have been largely studied during the last decades, including those between polymers, polymers at interfaces and scrambles polymer-cells-bacteria.(3, 6-10)

To our knowledge, despite the great deal of basic and applied information in literature about biological colonization, there is not any realistic analysis about the first necessary process of protein migration, previous to bacteria colonization.

In this paper we analyze several possible mechanisms for CBP formation, rejecting those impossible or out of range. We also propose a physical model describing how blood penetrates into the catheter, which captures the basic physical phenomenology under clinical practice conditions.

**The main ideas**

Two fluids have to be considered: a perfusion flow injected into the venous system trough a catheter, composed by water and medical drugs and the blood at certain pressure trying to rise against the main perfusion stream.

In a first approximation, the perfusion fluid dominates the general hydrodynamics at the catheter tip since blood pressure into veins is not enough to overcome the perfusion pressure. However, there are several mechanisms that could cause at least, partial bood rising against the perfusion flow. Capillarity due to surface tension could force blood to ascend the catheter. This is only possible if the blood surface tension coefficient largely overcomes that of the perfusion fluid. It never happens in our case since both coefficients are of the same order of magnitude and the difference is not enough to surpass drag forces over the blood.

It is well known that a fluid flowing through a region bounded by walls exhibits a parabolic velocity profile. The fluid velocity just on the walls is zero (or very low) while it reaches a maximum value at the center. This velocity distribution is only true under laminar flow for which, dissipative forces dominate over kinematic ones. The Reynolds number $= vR/v$ accounts for this feature, where $v$, $v$ and $R$ are the kinematic viscosity, velocity and the tube radius, respectively. The flow becomes turbulent for Reynolds number > 1000 and the velocity distribution turns disordered. In the present case, a small lumen section, a low injection rate and the value for kinematic viscosity of the perfusion fluid (water), warrant a laminar flow over all working real conditions. The Reynolds number for the catheter and clinical conditions employed in this study is about few tenths. Thus a laminar Poiseuille flow is warranted and phenomena as fluid entrainment, produced by a boundary layer separation and/or turbulent diffusion are not expected.

Due to the parabolic profile of the flow, the velocities in the region near the wall are very low and, consequently, blood can diffuse there without too much difficulty, whereas near the center of the cross section the relative high velocities makes more difficult the upstream diffusion process. The extent to which this happens will be examined in the next sections.

In fact, this phenomenon of diffusion under the influence of a field of velocities has been under study for a long time since the pioneering works of G.I. Taylor and other authors, and is known as "Taylor Dispersion", where the advection enhances diffusion, leading to an effective diffusion characterized by a coefficient that increases with the square of the mean velocity.(11-13)

Different to that, our model complements to some extent that viewpoint with the consideration of counterflow concentration variations. This makes the problem new and appealing since, as far as we know, no description has been made of the concentration profile on this setup.

We propose the "flow limited diffusion" as responsible for CBP. The velocity of the perfusion flow is very low close to the wall and the blood can diffuse counter-stream slipping on the wall. Far from the wall a field of relative high velocities frustrates the upstream blood diffusion, avoiding CBP. This is the basic idea we defend and implement as a physical model.

The flow limited diffusion depends on the intrinsic diffusion of blood along the perfusion fluid and on the motion of that fluid, influenced by viscosity, pressure gradients and geometry. Since the problem involves a competition between diffusion and advection, the solution must depend strongly on the Péclet number:

$$Pe = \frac{R^3 \nabla P}{D\eta} = \frac{\tau_D}{\tau} \qquad (1)$$

where $D$ is the diffusion coefficient of blood into the perfusion fluid, $\nabla P$ is the pressure gradient (variation of fluid pressure along the catheter length) and $\eta$ is the perfusion fluid dynamic viscosity. This adimensional number is easily interpreted as a competition between the time scale corresponding to the molecular diffusion, $\tau_D = R^2/D$ and the characteristic time for the flowing fluid $\tau = \eta/R\Delta P$. The Péclet number for the catheter in this paper is of the order of $10^5$, showing that diffusion is many times slower than advection.

**Bio-deposit Formation**

***Governing equation:*** The objective of the present work is to determine the physical mechanism responsible for the transport of biological material to form the biofilm observed into used medical catheters. Since an accumulation of biological material is observed along the catheter in the longitudinal direction, we assume that the corresponding diffusion transport should be proportional to the gradient of concentration (described in terms of the mass fraction $y$):

$$\mathbf{j} = -\rho D \overrightarrow{\nabla} y, \qquad\qquad\qquad (2)$$

$D$ being the Fick diffusion coefficient.

The first point is to check whether this description predicts a strong enough diffusion transport to counteract the convective transport in the opposite direction, with intensity proportional to the flow velocity. Assuming that molecular diffusion and convective mass transport are the only two relevant processes in the bulk fluid, the mass fraction of biological material is given by the solution of the Partial Differential Equation (PDE)

$$\rho \left( \frac{\partial y}{\partial t} + v \frac{\partial y}{\partial z} \right) = \frac{1}{r} \frac{\partial}{\partial r} \left( r \rho D \frac{\partial y}{\partial r} \right) + \frac{\partial}{\partial z} \left( \rho D \frac{\partial y}{\partial z} \right), \qquad (3)$$

where the catheter is a cylinder of radius $R$, $z$ is the longitudinal coordinate along the cylinder (with $z = 0$ at the catheter base) and $r$ is the radial coordinate. Axial symmetry has been assumed and consequently no azimutal dependence is expected. $v$ is the velocity of the fluid along the catheter for a Poiseuille flow:

$$v = \frac{-1}{4\eta} \left| \frac{\Delta p}{\Delta z} \right| \left( R^2 - r^2 \right). \qquad\qquad\qquad (4)$$

The LHS of Equation (3) is the convective derivative, which incorporates the advection, the RHS corresponds to diffusion. Equation (3) describes the behaviour of the concentration of biological material under the concurrence of both advection and diffusion phenomena.

***Dimensionless Variables:*** The catheter radius $R$ defines a characteristic length as well as the Fick diffusion coefficient sets a characteristic time $t_c = R^2/D$. Using these two units for distance and time and assuming constant density, the governing equation is written as:

$$\frac{\partial y}{\partial t} - 2Pe \left( 1 - r^2 \right) \frac{\partial y}{\partial z} = \frac{1}{r} \frac{\partial}{\partial r} \left( r \frac{\partial y}{\partial r} \right) + \frac{\partial^2 y}{\partial z^2}. \qquad (5)$$

The Péclet number accounting for the balance between convective and diffusive transport can be expressed as:

$P_e = RU/D$, where $U$ is the average fluid velocity $U = \frac{1}{8\eta} \left| \frac{\Delta p}{\Delta z} \right| R^2$. Since the Péclet number in our case is roughly $10^6$, global convective transport largely dominates over upward diffusive transport for the length scale $R$ and the diffusive time scale $R^2/D$ This means that a local description of solute transport must be made to explain the presence of biological material at a distance at least of the order of $R$.

***Why blood penetrates the catheter. Initial transient:*** The diffusive transport is proportional to the second derivative, whereas convective transport is proportional to the first derivative. This means that there exist a short enough length scale in which both phenomena are equally important, i.e., when both terms, the convective and diffusive, in Equation (5) are of the same order of magnitude.

The initial condition for the problem shows a strong discontinuity for *Y* at *z = 0* (*Y(t=0, z=0) = 1, Y(t=0, z>0) = 0*, independent of *r*). This means that there will be an intense diffusive transport towards the direction of coordinate *z* at *t=0*. The length scale, L, for which the diffusive transport becomes apparent, can be estimated taken into account that both convection and diffusion have to be of the same order of magnitude. This condition leads to *L = R/(2P$_e$)* at the axis of the catheter. For the present case, it is an extremely small fraction of *R*. The time scale (diffusive transient, t$_i$) on which the upward diffusion will be observable may be also determined in a similar way, leading to t$_i$ = L$^2$/D, imposing that the order of magnitude of the initial transient is such that the time derivative shows the same order of magnitude as the convective term.

Summarizing, during the initial transient time t$_i$, there will be a significant diffusive transport of biological material against the perfusion flow, reaching distances of the order of R/2Pe. For time larger than the time scale t$_c$, the convective transport largely dominates over the diffusive, trying to transport biological material out of the catheter.

As a consequence of the Poiseuille velocity profile, the diffusive transient lasts longer for the region close to the catheter surface than in the central area, since the velocity vanishes at $r = R$. The penetration length calculated at a distance $r$ from the centre is $L = \dfrac{R}{2Pe(1-r^2)}$ and the diffusive transient time $t_i = \dfrac{R^2}{4Pe^2(1-r^2)^2}$. This indicates that the transient dominates near the wall and the upstream diffusion will dominate for longer times, reaching longer distances travelling through a thin film very close to the wall. This is the reason why the biodeposit grows forming very thin layers of several microns on the internal wall. **Figure 1** plots the penetration length against r. Analogous behaviour can be observed for $t_i$.

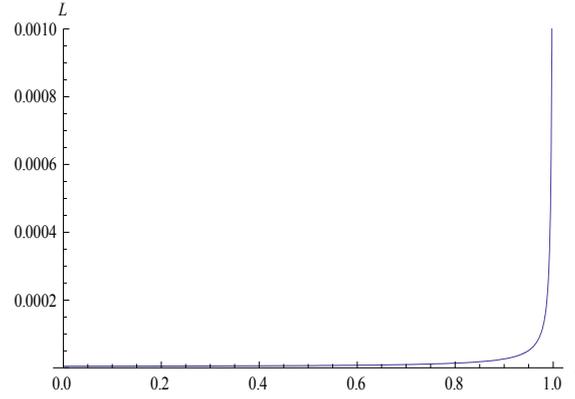

**Figure 1.** Dependence of the penetration length with r for $P_e=10^5$. L and r are in units of R.

This result shows that the particular problem conditions set the existence of a diffusion corridor in which blood diffuses "counter stream". The closer is to the wall, the longer its lifetime.

At this point, we can address a practical basic recommendation to delay the formation of biodeposit in medical practice. It is important to replace the perfusion flasks without any flow discontinuity, keeping a high injection speed.

**A thin diffusive corridor**

In order to estimate the thickness of the diffusive layer $\delta R$ we consider the problem at a radius given by $r = R(1 - \delta)$ and impose the condition that both terms of the diffusion transport are as important as the convective transport. This condition leads to $\delta = (4Pe)^{-1/2}$, which is a very small number. Thus, for large Péclet numbers, diffusion against the imposed stream restricts to an initial transient and depends on the distance to the wall, being the width of that layer inversely proportional to the square root of the Péclet number.

Another way to deal with this result is to evaluate directly the distance to the wall to which the diffusion process compares to advection. The comparison is performed by determining the region where both time scales are of the same magnitude. As the velocity profile depends on the distance to the wall, the characteristic time for advection depends also on the distance to the wall, conferring locality to the mechanism. Additionally, diffusion exhibits symmetry of concentric cylinders, locus of equal velocity.

A characteristic time for local advection may be defined as the ratio of a characteristic length λ along the z-direction and the local velocity:

$$\tau = \frac{\lambda}{\frac{\nabla P}{4\eta}\left(R^2 - r^2\right)} \qquad (6)$$

Similarly, the characteristic time for diffusion is $\tau_D = \lambda^2/D$. If $\lambda = 2\delta R$ is chosen as length due to diffusion isotropy, the comparison of both times in a region close to the wall, i.e., when $R\text{-}r = \delta R$, with $\delta \ll 1$, gives $\delta = \left[4Pe\right]^{-\frac{1}{2}}$, just as before.

In order to study the behavior of the system in such a thin diffusive layer for very large Péclet numbers, an asymptotic expansion is performed, introducing the stretched dimensionless coordinates X, Z and T, according to: $X = 2P_e^{\frac{1}{2}}(1\text{-}r)$, $Z = 2P_e^{\frac{1}{2}}z$, $T = 4P_e t$, where the position of the catheter surface is given by $X = 0$, while $X \rightarrow \infty$ represents the limit of the diffusive layer at the limit $P_e \rightarrow \infty$. The governing equation for the dimensionless coordinates is:

$$\frac{\partial y}{\partial T} - X\frac{\partial y}{\partial Z} = \frac{\partial^2 y}{\partial X^2} + \frac{\partial^2 y}{\partial Z^2}, \qquad (7)$$

strictly valid for the thin diffusive layer and infinite Péclet number.

While the downwards convective transport vanishes at the catheter surface ($X = 0$), the Fick diffusion in the radial direction tends to push the material away from the catheter's surface, into a region where convection simply washes out the solute. We consider two possibilities to explain the presence of biosubstrate: i) the biological material diffuses against the flow inside the thin diffusive layer, according to equation (7), and instantaneously attaches to the boundary, thus preventing being washed out by the flow, or ii) the biological material diffuses directly on the catheter surface, by surface diffusion. In both cases, CBP advances along the catheter length as $\left(Dt\right)^{\frac{1}{2}}$. In the first case, $D$ is the diffusion coefficient corresponding to Fick diffusion in the liquid, whereas in the second case, $D$ is the diffusion coefficient corresponding to surface diffusion. It is quite unlikely that both coefficients should match. Actually, surface diffusion is many orders of magnitude slower.

In this work we deal with the first mechanism, since surface diffusion is probably too slow in the dimensions and curvatures determined by the catheter geometry.

**Instantaneous solute deposition**

Blood is a complex fluid where many components and interactions are present. Thus, fibrin deposition on

catheter´s wall is unlikely to be due to simple deposition. Nevertheless, in our opinion, a simple model of deposition could illustrate about the characteristics and the spatial and temporal scales involved in the problem. In this respect, we propose the following description: As a portion of solute diffusing up in $z$ direction will also diffuse towards $X$ direction, away from the catheter surface, reaches a region where the Poiseuille flow tends to wash it out. Other portion of solute will diffuse towards $X=0$ to form a solid deposit on the surface, as observed. Once the solute deposits on the catheter surface, it acts as a sink of biological material

The simplest way to model the process of solute deposition is to assume that it happens instantaneously (*i.e.* in a characteristic time scale much shorter than any other time scale of the problem). The boundary condition imposed on the catheter surface is Y=0 at $X=0$. Let us recall that, even though this boundary condition is applied at $X=0$, it refers to the liquid phase. This boundary condition means that, as soon as any solute touches the catheter surface it deposits there, leaving the liquid phase.

**Rate of Deposition**

Prior to obtain the solution to the governing equation under this new boundary condition, let us revisit the concept of the mass of the solid deposit. The mass flow towards the catheter surface (mass per unit time and per unit surface) is given by the Fick law:

$$j = -\rho D \frac{\partial y}{\partial r} \quad \text{at} \ \ r=R \ , \qquad (8)$$

written in dimensionless units as:

$$j = \frac{-\rho D}{R} \frac{\partial y}{\partial r} \ \text{at} \ r=1 \ . \qquad (9)$$

Accordingly the variation of mass, $m$, per surface unit is given by

$$\frac{dm}{surface} = j dt = -\rho R \frac{\partial y}{\partial r}(r=1) dt \ , \qquad (10)$$

$t$ being the time in units of the diffusion time, $R^2/D$.

From this relationship we see that $\rho R$ is the natural scale for surface mass density, accordingly, using this quantity as the scale for dm/surface:

$$dm = \frac{-\partial y}{\partial r}(r=1) dt \rightarrow m = -\int \frac{\partial y}{\partial r}(r=1) dt \ . \qquad (11)$$

It can be written in terms of the scaled variables of the thin diffusive layer as:

$$\cdot \, dm = \frac{1}{2\sqrt{Pe}} \frac{\partial y}{\partial X}(X=0)dT \rightarrow m = \frac{1}{2\sqrt{Pe}} \int \frac{\partial y}{\partial X}(X=0)dT \quad (12)$$

**Analytical Solution for the Rate of Deposition**

In the present section it will be shown that the boundary condition of instantaneous solute deposition on the catheter's surface allows for a simple analytical solution of the deposition rate. To that end, let us assume that in the thin diffusive layer the mass fraction $y$ is an analytic function of $X$. Hence, for very small $X$ values, $y$ is given by the first order Taylor expansion $y(X, Z, T)=y(X=0, Z, T)+J(Z,T)X$, where $J(Z,T)=\partial y/\partial X$ at $X=0$. This relationship is written as: $y(X, Z, T)=J(Z,T)X$, once the boundary condition of instantaneous solute deposition introduced. The derivative of the governing equation (7) at the thin diffusive layer is then taken and evaluated at $X = 0$. Finally, a closed relationship for the deposition rate $J(Z,T)$ is found:

$$\frac{\partial J}{\partial T} = \frac{\partial^2 J}{\partial Z^2} \quad (13)$$

In order to solve this equation, the corresponding boundary and initial conditions are imposed. It is clear that the deposition rate has to vanish for large enough $Z$, and $J(Z\rightarrow\infty, T) = 0$ verifies. In addition, as we approximate to the catheter base ($Z=0$) the deposition rate must become independent of $Z$, and $\partial J/\partial Z=0$ at $Z=0$.

Finally, for the initial conditions we choose $J(Z>0, T=0) = 0$, whereas $J(Z=0, T=0) = \infty$. The reason for this initial condition is a consequence of the approximation of instantaneous solute deposition, together with the fact that the catheter surface (at $Z = 0$) is in contact with a non vanishing solute mass fraction at $t = 0$, yielding an infinity rate of deposition at $Z = 0$. In other words, catheter´s tip is initially a frontier of discontinuity for solute concentration, where just out of it the solute exists whereas just inside it, there is no solute. The strong initial discontinuity in solute´s gradient can be expressed though he rate of deposition, *i.e.* $J(T=0)=\delta(z)$.

The former simplified model shows a straightforward analytical solution for the rate of deposition:

$$J = \frac{\exp\left(-Z^2/(4T)\right)}{\sqrt{(4\pi T)}} \quad (14)$$

This information in introduced into the time integral giving as output the mass of solid deposit as a function of $Z$ and $T$:

$$m = \frac{1}{2\sqrt{Pe}} \int J(Z,T)dT \quad (15)$$

Finally, the mass (per unit surface) of solid deposit (in units $\rho R$) as a function of $Z$ is:

$$m(Z,T) = \frac{1}{2\sqrt{Pe}}\left[\sqrt{\frac{T}{\pi}}e^{-(Z^2/4T)} - \frac{Z}{2}Erfc\frac{Z}{2\sqrt{T}}\right] \quad (16)$$

This analytical solution contributes to the explanation of the observed biofilm substrate. Hence, the physical nature of the phenomenon is the Fick diffusion in an extremely thin diffusion layer next to the catheter's surface, followed by instantaneous deposition on the catheter as soon as any biological particle reaches it.

The first comment to point out is that diffusion always occurs, no matter the values of the chosen parameters. It means that CBP is unavoidable. Besides, diffusion occurs only close to the wall. The larger the value of $Pe$, the closer to the wall the diffusion occurs. This finding is in agreement with the fact that when drag increases, the blood diffusion upstream (following the axis of the catheter) is hindered by the flow. However, since the flow velocity is zero at the wall, the diffusion occurs only very close to it. This makes the CBP process equivalent to a one-dimensional diffusion problem in which, the diffusion front moves as $t^{\frac{1}{2}}$ so that the time needed for CBP to attain a given height inside the catheter becomes predictable.

**Conclusions**:

The model presented above complements to some extent the problem of Taylor Dispersion, when a solute diffuses into a stream. In our case, the formulation of flow limited diffusion of blood inside endovenous catheters in clinic conditions of intensive care units, it was elucidated that a very thin diffusion layer very close to the wall of the catheter is determinant to produce CBP with the consequent fibrine deposition, the value of the Péclet number playing a crucial role in the width of the layer. This clarifies the origin of biofilm formation in these cateters.

The elucidation of the physical mechanism responsible for the formation of the biofilm is essential in order to ellaborate adequate protocols for endovenous catheter handling, perfusion regimes and administration.

**Acknowledgment:**

This work was supported by Gobierno de España - Ministerio de Ciencia e Innovación under project MAT2009-14234-C03-02, Junta de Andalucía under "Excellence Project": FQM-02353 and European Union under project "Colloidal and Interface Chemistry for Nanotechnology" (COST Action - 43). S.M.E thanks to the Instituto Carlos Tercero (Ministerio de Ciencia e Innovación. Spain) for the fellowship which allows a long period research at the Group of Complex Fluid Physics and Nanotechnology laboratory (University of Almería).